\documentclass[twocolumn]{pasj00}

\begin{document}
\SetRunningHead{Y. Takeda and M. Takada-Hidai}{Chromospheres in Metal-poor Stars}
\Received{2010/11/05}
\Accepted{2010/12/24}

\title{Chromospheres in Metal-Poor Stars\\
Evidenced from the He~I 10830~$\rm\AA$ Line
\thanks{Based on data collected at Subaru Telescope, which is operated by 
the National Astronomical Observatory of Japan.}
}

%

\author{Yoichi \textsc{Takeda}}
\affil{National Astronomical Observatory of Japan 
2-21-1 Osawa, Mitaka, Tokyo 181-8588}
\email{takeda.yoichi@nao.ac.jp}
\and 
\author{Masahide \textsc{Takada-Hidai}}
\affil{Liberal Arts Education Center, Tokai University, 
4-1-1 Kitakaname, Hiratsuka, Kanagawa 259-1292}
\email{hidai@apus.rh.u-tokai.ac.jp}

%


\maketitle

\begin{abstract}

Based on the near-IR spectra of 33 late-type stars in the wide 
metallicity range (mainly dwarfs and partly giants) obtained 
with IRCS+AO188 of the Subaru Telescope,
we confirmed that He~{\sc i} 10830~$\rm\AA$ line is seen in absorption
in almost all moderately to extremely metal-poor stars of thick disk and 
halo population (from [Fe/H]~$\sim -0.5$ down to [Fe/H]~$\sim -3.7$), 
the strength of which is almost constant irrespective of the metallicity. 
This is an evidence that chromospheric activity at a basal level 
persists even for such old stars, despite that their rotations are 
considered to be slowed down and incapable of sustaining a dynamo, 
suggesting that some kind of chromospheric heating mechanism independent 
of rotation/magnetism (e.g., acoustic heating) may take place.

\end{abstract}

%


\section{Introduction}

It is not well understood whether and how the chromosphere (temperature 
rise phenomenon in the upper atmosphere), which is confirmed in the Sun 
and nearby solar-type stars, exists in old metal-poor stars.

A useful guideline for considering this problem is the 
positive correlation between the chromospheric emission
and surface rotation among cool (young to solar age) stars 
(Noyes et al. 1984), which can be rephrased as the activity--age
relation (Skumanich law), since rotation is generally decelerated 
with age as a result of magnetic braking (i.e., loss of angular
momentum caused by stellar wind pulling out magnetic field lines).
This observational fact is generally interpreted by the scenario
that globally-organized magnetic field generated by
differential rotation-induced dynamo is essentially responsible
for the origin of the chromospheric activity.
Then, we may expect that any chromospheric activity would have 
almost come to rest in such aged stars, as they are likely to be rotating
very slowly (i.e., practically non-rotating) due to the long-lasting
gradual deceleration they have suffered. 

Meanwhile, contrary to this naive expectation, there are 
indications that chromospheres {\it do} exist even in very old 
population~II stars. 
For example, Dupree, Smith, and Strader (2009) recently showed, in their 
extensive studies on the infrared He~{\sc i} line at 10830~$\rm\AA$ 
(hereinafter often abbreviated as ``He~{\sc i} 10830'') for 
41 metal-deficient G--K giants, that this high-excitation line arising 
from the metastable level (2s$^{3}S_{1}$, $\chi_{\rm low}$ = 19.72~eV) 
is ubiquitously observed with appreciable blue shift for stars with
$T_{\rm eff} \gtsim 4500$~K, indicating the existence of 
outward-moving high-temperature chromospheric layer.
Actually, detections of chromospheric emission in the core of 
Mg~{\sc ii} $h+k$ doublet at 2800~$\rm\AA$ were also reported 
for 10 population~II giants by Dupree, Hartmann, and Smith (1990).

Yet, it is not clear from this fact alone whether the chromospheric
heating mechanism similar to the solar case is operative in these 
evolved red giants, since some other physical process 
specific to the unstable low-density atmosphere may be 
responsible for this activity (for example, pulsation-induced 
shock dissipation; e.g., Smith \& Dupree 1988).
Therefore, in order to see how the solar-type chromospheres
go through changes towards metal-poor regime, one should study
the activity of unevolved population~II ``dwarfs.''

As to activities of metal-poor dwarfs, we know only a few researches.
Smith and Churchill (1998) investigated the cores of Ca~{\sc ii} H+K 
lines in 23 metal-poor dwarfs, and found signs of chromospheric emission 
in some halo stars with metallicities as low as [Fe/H]~$\sim -2$, 
while negative results were also obtained for not a few cases. 
Peterson and Schrijver (1997) detected chromospheric emissions
of Mg~{\sc ii} $h+k$ 2800 lines (being superior to Ca~{\sc ii} H+K 
lines for detecting chromospheric core emissions, because of the 
difference in the line- as well as continuum-opacities) in all 
($\sim 10$) metal-poor solar-type stars down to 
[Fe/H]~$\sim -2.60$ they searched, which indicates that 
chromospheres ubiquitously exist in population~II halo dwarfs. 
Peterson and Schrijver (2001) further confirmed the existence of 
Lyman~$\alpha$ emission in their sample of halo stars (though
the contamination of interstellar absorption and geocoronal 
emission made any detailed line-profile study rather difficult), 
which substantiates their conclusion.

Though these studies invoking the emission in the core of strong 
metal-lines (Ca~{\sc ii} H+K or Mg~{\sc ii} $h+k$) are regarded 
as important, since the existence of chromospheres in old 
population~II dwarfs has been revealed in the qualitative sense, 
they are not necessarily informative when it comes to 
quantitatively studying whether or not the degree of activity 
varies with the metallicity, since the extent of core emission 
in such metallic lines would depend on the metallicity even if 
the extent of activity remains unchanged; also, its usability is not 
clear for extremely metal-poor regime (e.g., [Fe/H]~$\ltsim -3$). 

From this viewpoint, the most desirable way to investigate the 
chromospheric activity of metal-poor solar-type stars is to use 
the He~{\sc i} 10830 line, since it is a chromospheric indicator
practically independent of the metallicity.
Admittedly, several extensive studies on the He~{\sc i} 10830 line 
for solar-type stars have been published so far
(e.g, Vaughan \& Zirin 1968; Zirin 1982; Zarro \& Zirin 1986;
O'Brien \& Lambert 1986; Lambert 1987; see also the references
cited in Andretta \& Giampapa 1995). Somewhat unexpectedly,
however, any observation of the He~{\sc i} 10830 line seems to 
have never been reported for ``metal-poor dwarfs of population~II'' 
to our knowledge.

Timely, we have recently obtained near-IR ($zJ$-band) spectra 
of 33 disk/halo stars (mainly dwarfs and partly giants) 
covering a wide metallicity range ($-3.7 \ltsim$~[Fe/H]~$\ltsim +0.3$),
and carried out sulfur abundance determinations based on
the S~{\sc i} triplet lines at 10455--10459~$\rm\AA$
(Takeda \& Takada-Hidai 2011; hereinafter referred to as Paper I).
Therefore, we decided to examine the He~{\sc i} 10830 line in the 
spectra of these stars with an aim to see how the chromospheric 
activity behaves itself in metal-poor stars,
while checking if this line is visible and how its strength varies 
from star to star with a change of metallicity. This is the purpose 
of this investigation.

\section{Observational Data}

The list of the program stars, 33 halo/disk stars (24 dwarfs
and 9 giants\footnote{
As in Paper I, the targets are divided according 
to the surface gravity into two classes (those with $\log g < 3$ 
and $\log g > 3$), which we tentatively call ``giants'' 
and ``dwarfs,'' respectively. It should thus be kept in mind
that ``dwarfs'' includes some stars actually classified as 
subgiants. Besides, in this paper, giants with $\log g <2$ 
are particularly called as ``low-gravity giants'' (cf. table 1),
since they tend to show anomalous He~{\sc i} 10830 line profiles.})
in the metallicity range of $-3.7 \ltsim$~[Fe/H]~$\ltsim +0.3$, is 
presented in table 1, where the literature values of the 
atmospheric parameters (the same as in Paper I), apparent $V$ 
magnitudes (from SIMBAD database), and 
$\log (f_{\rm x}/f_{\rm bol})$ values (ratio of the X-ray flux
to the total bolometric flux) available in ROSAT archival 
data (accessible from the HEASARC web site\footnote{
$\langle$http://heasarc.gsfc.nasa.gov/$\rangle$.}) are also given.
The observational data are the $zJ$-band (1.04--1.19~$\mu$m) spectra 
with $R\sim 20000$ (wavelength resolution) and S/N~$\sim$~100--200 
(signal-to-noise ratio; $\sim 300$ only for BD+44$^{\circ}$493), 
which were obtained on 2009 July 29 and 30 (UT) 
with the Infrared Camera and Spectrograph (IRCS) along with
the 188-element curvature-based adaptive optics system (AO188)
of the Subaru Telescope. See section 2 of Paper I for the details
of the observation and the data reduction procedure.

Since several appreciable telluric lines (due to H$_{2}$O vapor)
are located in the neighborhood of the He~{\sc i} 10830 line,
we removed them by dividing each stellar spectrum by that of
a rapid rotator ($\gamma$~Tri) with the help the task 
``{\tt telluric}'' in IRAF.\footnote{IRAF is distributed by the 
National Optical Astronomy Observatories, which is operated by 
the Association of Universities for Research in Astronomy, Inc. 
under cooperative agreement with the National Science Foundation.} 
This procedure turned out quite successful, as shown in figure 1.
The finally resulting spectra, which were used for measuring 
the He~{\sc i} 10830 line, are displayed in figure 2.
Regarding the Sun, we adopted the solar flux spectrum published
by Kurucz et al. (1984).

\setcounter{figure}{0}
\begin{figure}
  \begin{center}
    \FigureFile(80mm,80mm){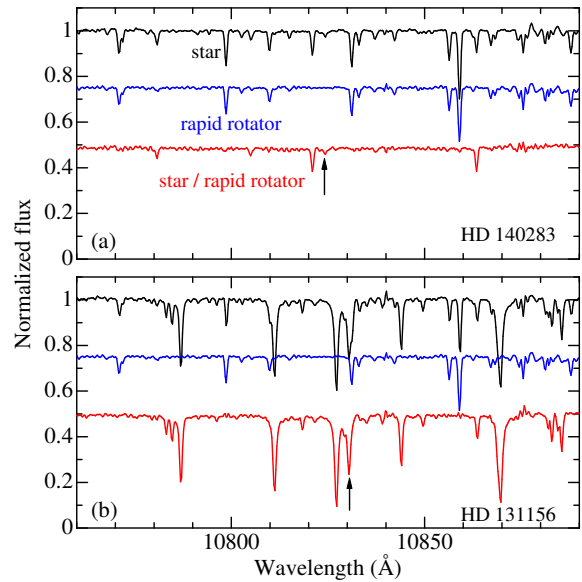}
  \end{center}
\caption{
Examples demonstrating how the telluric lines (mostly due to H$_{2}$O
 vapor) could be removed in the spectra used for measuring the He~{\sc i} 
10830 line. Dividing the raw stellar spectrum (top, black) by the 
spectrum of a rapid rotator $\gamma$~Tri (middle, with an offset 
of $-0.25$; blue) results in the final spectrum (bottom, with an 
offset of $-0.5$; red). 
The upper panel (a) and lower panel (b) show the typical metal-poor 
case (HD~140283) and the metal-rich  case (HD~131156), respectively. 
No Doppler correction is applied to the wavelength scale of these 
spectra. The arrow indicates the position of the He~{\sc i} 10830 line.
}
\end{figure}

\section{Measurement of the He~I 10830 Feature}

The He~{\sc i} 10830~$\rm\AA$ triplet 
(2s$^{3}S_{1}$---2p$^{3}P_{0,1,2}$) comprises three 
components at 10829.088, 10830.247, and 10830.336~$\rm\AA$ (with
$\log gf$ of $-0.745$, $-0.268$, and $-0.047$, respectively).\footnote{
Data from the Vienna Atomic Line Database 
$\langle$http://vald.astro.univie.ac.at/$\rangle$.}
However, since the first component is considerably weaker than 
and separated from the others, the latter two close components 
essentially determine the position of the stellar He~{\sc i} 10830 line.

First of all, the important fact manifestly recognized from figure 2 
is that this He~{\sc i} line is clearly visible (in absorption) in 
most of the program stars; i.e., not only metal-rich stars of 
disk population but also old very metal-poor stars (even in the extremely 
metal-poor star BD+44$^{\circ}$493 with [Fe/H]~$\simeq -3.7$). 
This is a reconfirmation of the previous report (based on Ca~{\sc ii}, 
Mg~{\sc ii}, and Ly~$\alpha$ lines) that chromospheres ubiquitously 
exist in old population~II solar-type stars (Smith \& Churchill 1998; 
Peterson \& Schrijver 1997, 2001); but our result substantiates 
this fact in a more convincing manner for a larger number of stars
in a much wider metallicity range than ever before.

The equivalent width ($EW$) of the He~{\sc i} 10830 line was evaluated
for each star by the Gaussian fitting, as depicted in figure 2.
We tried to measure $EW$ as long as (i) the center of the absorption
feature situates at almost the expected wavelength of the He~{\sc i} 
10830 line, and (ii) the line profile is not too severely blended.

Generally, ``low-gravity'' giants (with $\log g < 2$) tend to show 
appreciably blue-shifted components, making the 
profile complex or blended with the neighboring Si~{\sc i} line
at 10827.1~$\rm\AA$, which must be due to the significant gas outflow
typically seen in red-giant atmospheres (e.g., Dupree et al. 2009).
So we had to abandon measuring $EW$s for four low-gravity giants 
(HD~122563, HD~204543, HD~121135, and Arcturus).
As a result, since the $EW$ data for ``giants'' (with $\log g < 3$) 
turned out insufficient (only five stars), we will focus our 
discussion only on the 24 ``dwarf'' stars in the remainder of 
this paper (unless otherwise specially noted),
given that the main aim of this paper is to investigate the 
chromospheric activity of population~II ``solar-type'' stars. 

The line-depth ($R_{0}$), the full-width at half-maximum ($fwhm$), 
and the equivalent width ($EW$) corresponding to the fitted 
Gaussian profile are given in table 1, while figure 3 shows 
how the resulting $\log EW$ values are correlated with [Fe/H].

Uncertainties in $EW$ caused by photometric random errors may be 
estimated by invoking the formula derived by Cayrel (1988),
who showed that the ambiguity in $EW$ is roughly expressed as 
$\sim 1.6 (w\;\delta x)^{1/2} \epsilon$, where $w$ is the typical 
line FWHM, $\delta x$ is the pixel size (in unit of wavelength), and
$\epsilon$ is the photometric accuracy represented by 
$\sim$~(S/N)$^{-1}$. 
Substituting $w \sim 1$~$\rm\AA$, $\delta x \simeq 0.25 \rm\AA$, 
and $\epsilon \sim 1/100$, we obtain $\sim $~8~m$\rm\AA$ as 
the uncertainty in $EW$. Considering that typical $EW$ is 
$\sim 30$~m$\rm\AA$ in the present case, corresponding to
$\sim \pm$~0.1~dex in $\log EW$. Meanwhile, since the continuum level
could be reasonably defined to $\ltsim$~10--20\% of the line-depth 
for most of the cases (cf. figure 2), the errors in $\log EW$ 
due to uncertainties in the continuum placement are estimated to 
be again $\ltsim$~0.1~dex. Accordingly, combining these two factors,
we may state that the ambiguity in $\log EW$ is on the order of 
$\sim \pm$~0.1~dex ($\sim$~20--30\% in $EW$).

\setcounter{figure}{2}
\begin{figure}
  \begin{center}
    \FigureFile(80mm,80mm){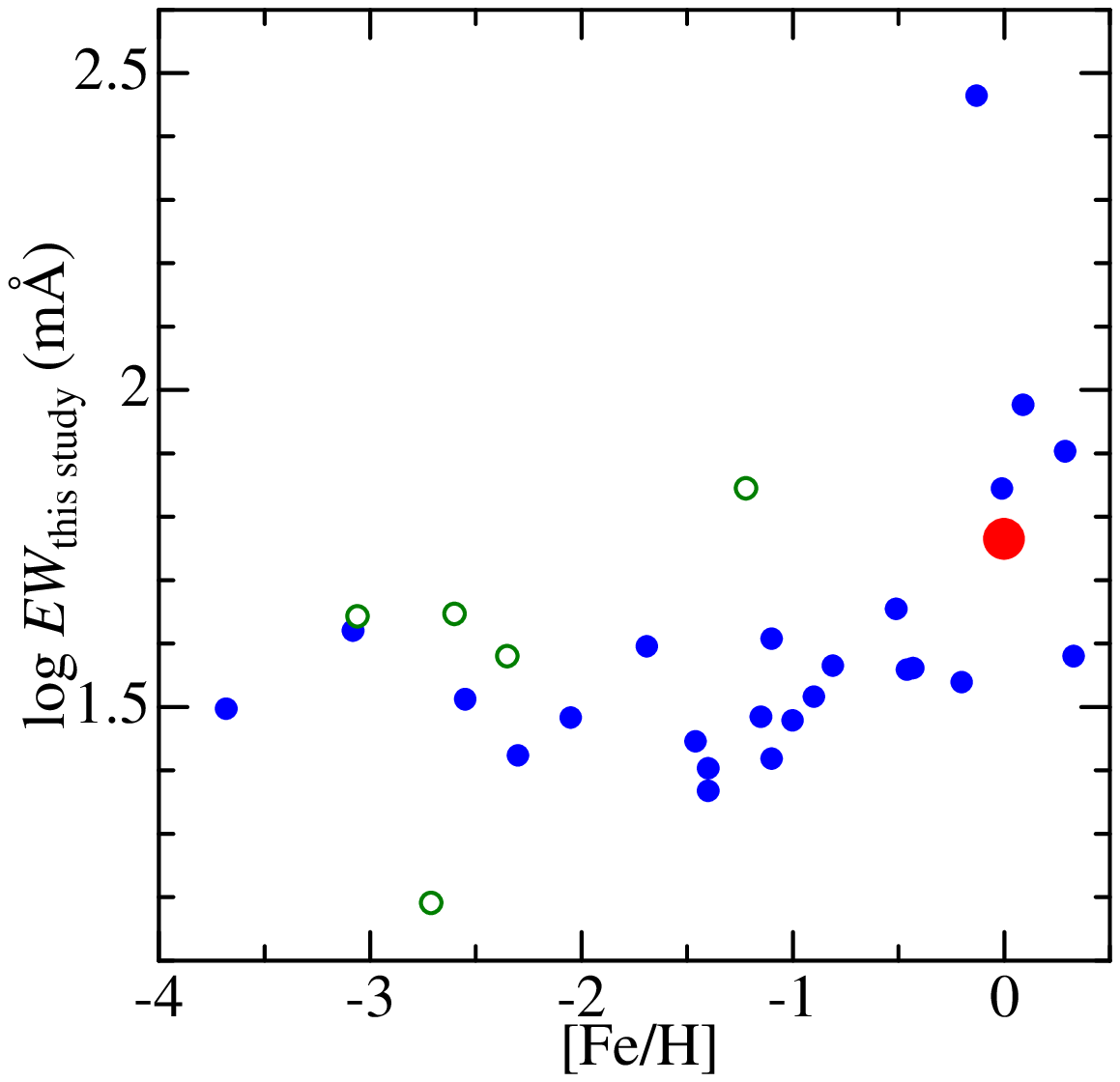}
  \end{center}
\caption{Logarithmic equivalent width ($\log EW$) measured for 
the He~{\sc i} 10830 line, plotted against the metallicity ([Fe/H]). 
The results for 24 dwarfs ($\log g > 3$) and 9 giants ($\log g < 3$) 
are discriminated by filled (blue) and open (green) symbols, 
respectively. The Sun is indicated by the large (red) filled circle.
}
\end{figure}

\section{Discussion}

\subsection{Characteristics in the Behaviors of EW(He I 10830)}

An inspection of figure 3 reveals the following notable chracteristics.
While we can not say much about giants (open circles) because 
of their paucity, which tend to show a rather large spread between 
$\log EW ({\rm m\AA}) \sim$~1 ad $\sim $~2, dwarfs (filled circles)
exhibit a significant trend in terms of the metallicity. 
That is, its strength is nearly constant around 
$\log EW ({\rm m\AA}) \sim 1.5$ over the wide 
metallicity range of $-3.7 \ltsim$~[Fe/H]~$\ltsim -0.5$, while 
it tends to somewhat strengthen (in the average sense) up to 
$\log EW({\rm m\AA}) \sim 2$ for comparatively metal-rich stars 
at $-0.5 \ltsim$~[Fe/H] 
(HD~131156 = $\xi$~Boo~A shows an exceptionally large strength 
of $\log EW ({\rm m\AA}) \sim 2.5$, which is a well-known active star 
with large spots on its surface as characterized by the detection 
of its magnetic field by Robinson et al. 1980).  

For the purpose of corroborating this relation, we increased 
the data poits by adding the literature results taken from 
Zarro and Zirin (1986) and Dupree et al. (2009) as displayed 
separately for dwarfs (figure 4a) and giants (figure 4b).
Considering the large variation of $EW$(dwarfs) in the metal-rich 
domain around [Fe/H]~$\sim 0$ as manifestly seen in figure 4a, 
we may regard that such an apparent increase in the (averaged) 
$EW$(dwarfs) at $-0.5 \ltsim$~[Fe/H] seen in our data (figure 3) 
simply reflects the difference in the dispersion of $EW$(dwarfs) 
between the [Fe/H]~$\ltsim -0.5$ group (small dispersion) 
and $-0.5 \ltsim$~[Fe/H] group (considerably large dispersion). 
It may be worth stressing that such a near-constancy in $EW$ 
for metal-poor stars at [Fe/H]~$\ltsim -0.5$ is restricted 
to ``dwarf'' stars, since $EW$(giants) show a considerably large
scatter irrespective of the metallicity (figure 4b). 

Further, we plotted the measured $EW$ against  
$\log (f_{\rm x}/f_{\rm bol})$ observed by ROSAT 
(available only 8 apparently bright dwarf stars of $V \ltsim 5$
at $-0.9 \ltsim$~[Fe/H]; cf. table 1) in figure 5.
We can see from this figure that our data reasonably follow 
the tendency exhibited by the various published data compiled 
by Zarro and Zirin (1986). 

\setcounter{figure}{3}
\begin{figure}
  \begin{center}
    \FigureFile(80mm,80mm){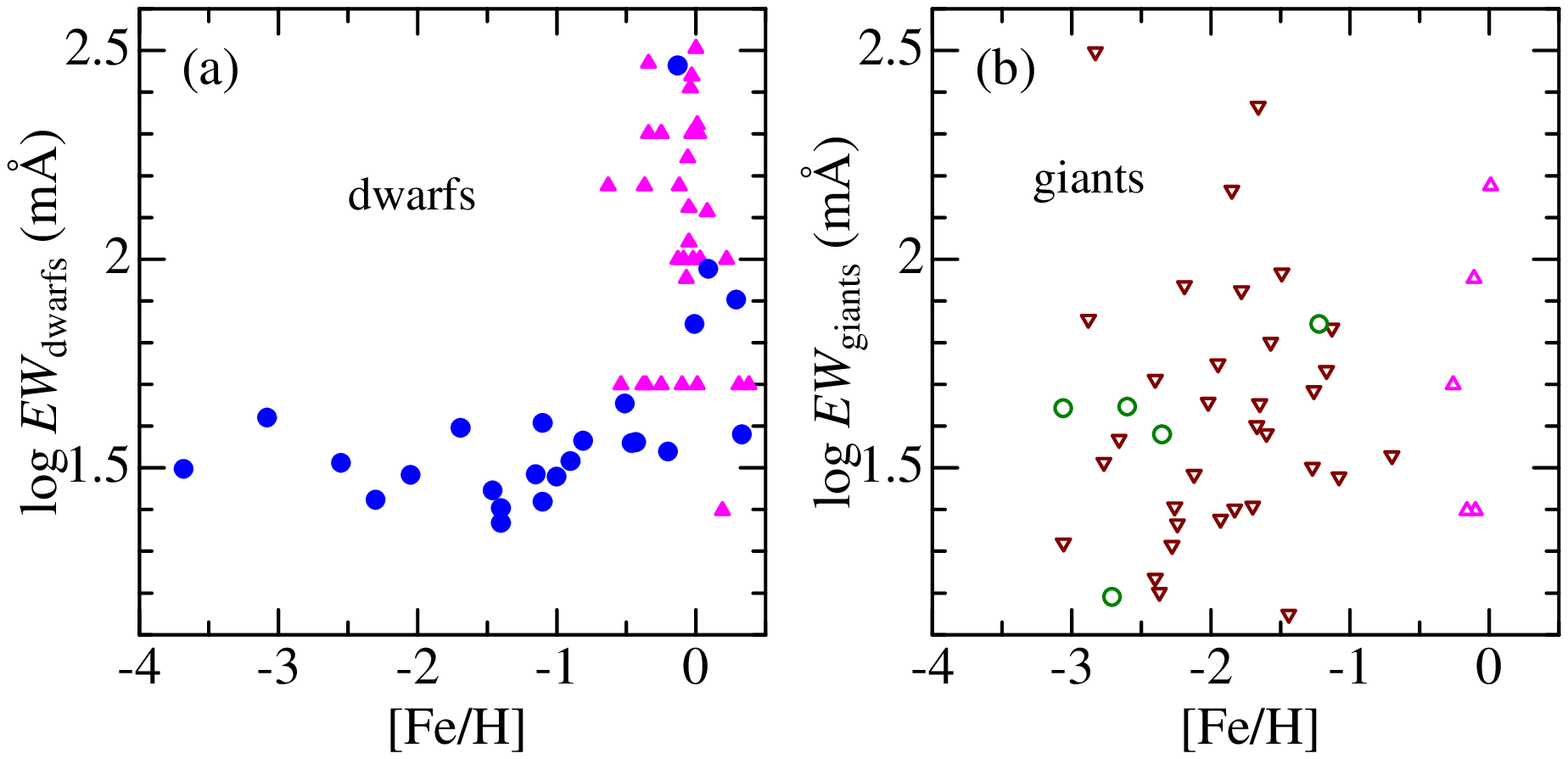}
  \end{center}
\caption{Comparison of the $\log EW$(10830) vs. [Fe/H] relation
derived from this study with that from the literature data. 
Panels (a) and (b) correpond to dwarfs (filled symbols) and 
giants (open symbols), respectively.
Circles $\ldots$ this study (same colors as in figure 3); 
triangles (pink) $\ldots$ Zarro and Zirin's (1986) compilation 
($EW$) along with Cayrel de Strobel's (2001) catalogue ([Fe/H]);
inverse triangles (brown) $\ldots$ Dupree et al. (2009).
}
\end{figure}

\setcounter{figure}{4}
\begin{figure}
  \begin{center}
    \FigureFile(80mm,80mm){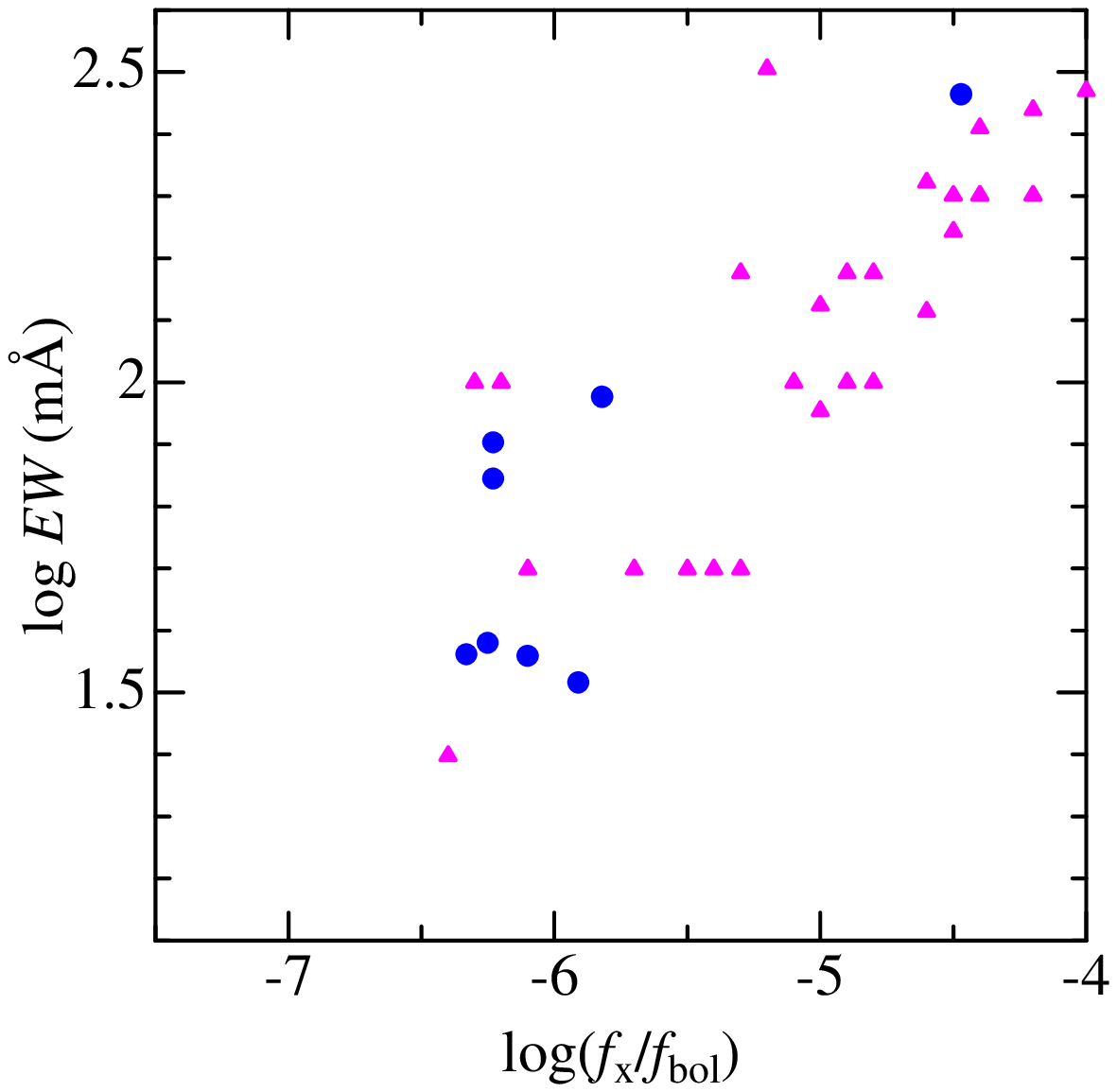}
  \end{center}
\caption{Logarithmic He~{\sc i} 10830 line strength ($\log EW$) 
plotted against X-to-bolometric flux ratio
($\log (f_{\rm x}/f_{\rm bol})$). Circles (blue) $\ldots$ $EW$ data 
derived in this study compared with the ROSAT data (cf.table 1);
triangles (pink) $\ldots$ published data compiled by Zarro and Zirin (1986). 
Note that all the data shown here are for dwarf stars, since 
available X-ray data are essentially limited to apparently 
bright dwarfs.
}
\end{figure}

\subsection{Origin of Chromospheres}

From the viewpoint of the kinematic properties, the two metallicity
groups (roughly divided at [Fe/H]~$-0.5$) mentioned in the previous
subsection are markedly different from each other (see figure 1 in 
Paper I); i.e., the former is ``halo or thick-disk stars'' (older than 
the Sun) while the latter is ``thin-disk stars'' (near-solar age 
or younger stars).
Considering the importance of rotation in the conventional
understanding of stellar activity described in section 1, we may 
speculate that different types of chromospheres are involved between 
these two cases:\\
---(1) For comparatively young thin-disk stars ($-0.5 \ltsim$~[Fe/H];
where a markedly large dispersion in $EW$ is observed),
which still show appreciable rotational velocities, stellar rotation 
plays the decisive role in controlling the chromospheric activity,
for which organized magnetic fields generated by a dynamo (which 
is induced by the differential rotation) are responsible.\\
---(2) For older halo or thick-disk stars
($-3.7 \ltsim$~[Fe/H]~$\ltsim -0.5$; where $EW$ is almost constant) 
whose rotations are supposed to have been considerably slowed down, 
the standard rotation/magnetism-induced heating mechanism does not 
work efficiently any more.
Instead, a chromospheric activity at the ``basal'' or minimum level 
(e.g., Schrijver 1987; Rutten et al. 1991) comes into sight,
which is maintained irrespective of stellar rotation.
Regarding the specific mechanism responsible for such a ``basal''
chromosphere, acoustic waves generated by convection (apparently
independent of rotation) may be promising (e.g., Schrijver 1995;
Narain \& Ulmschneider 1996; see also the discussion in
Peterson \& Schrijver 1997, 2001). 

We point out that, in order to test the ``basal chromosphere'' 
scenario mentioned above, long-term observations (over $\sim 10$ 
years or more) monitoring the activity of old metal-poor stars 
may be worth carrying out, since those stars would not show any 
near-cyclic activity variation such as that exhibited by the Sun 
or nearby solar-type stars. The He~{\sc i} 10830 line 
(a good chromospheric indicator as evidenced by its close connection
with the Ca~{\sc ii} H+K core emission in the Sun seen as a star; 
e.g., Livingston et al. 2007) would be 
most suitable for this purpose, as it can be used even for extremely 
metal-deficient stars (unlike Ca~{\sc ii} H+K) and its observation 
can be done from the ground (unlike Mg~{\sc ii} $h+k$ or Lyman~$\alpha$).

\subsection{Rotations of Metal-Poor Stars}

To make this discussion complete, we should remark that our 
argument is based on the (intuitively reasonable) postulation 
that rotational velocities of old metal-poor stars are appreciably 
slowed down compared with those of younger metal-rich stars.
It is thus important to check if such a tendency is really observed. 
But embarrassingly, according to the literature data compiled by
Cor\'{t}es et al. (2009), metal-deficient stars ([Fe/H] $\ltsim -1$) 
in our sample show even {\it higher} $v_{\rm e}\sin i$ values than 
metal-rich stars as shown in figure 6, which is just the opposite 
to our expectation.
However, considering the enormous difficulty involved with 
$v_{\rm e}\sin i$ determinations of slow rotators as low as 
$\ltsim $1--2~km~s$^{-1}$ (i.e., contribution of the macroturbulence 
is almost predominant over the rotational effect on the line profile), 
we are still conservative and rather reluctant to take these results
seriously, which are based on not-so-high quality spectra and subject 
to errors on the same order of the $v_{\rm e}\sin i$ values themselves
(cf. table 1 of Cor\'{t}es et al. 2009).\footnote{For example, by using 
the high-quality Subaru/HDS spectra of BD+44$^{\circ}493$ and 
HD~140283 available to us (visual region, $R \simeq 90000$, 
S/N~$\sim 350$ and $\sim 700$, respectively), 
we carried out $v_{\rm e}\sin i$ determinations for these stars, 
based on the profile fitting for Fe~{\sc i} 5586.77 and Ca~{\sc i} 
5588.75 lines following the procedure described in Takeda et al. 
(2010; cf. section 3 therein). 
We then obtained almost the same results of 
$v_{\rm e}\sin i \simeq$~2.3--2.4~km~s$^{-1}$ for both stars, 
which suggests that the $v_{\rm e}\sin i$ values of 3.9~km~s$^{-1}$ 
(BD+44$^{\circ}$493) and 5.0~km~s$^{-1}$ (HD~140283) given in table 3 
of Cor\'{t}es et al. (2009) are erroneously overestimated.} 
In any event, if it were really confirmed based on reliable 
observations that rotations of old stars are not decelerated 
(or even accelerated), then we would have to earnestly reconsider the 
stellar activity of metal-poor stars in a completely different paradigm. 

\setcounter{figure}{5}
\begin{figure}
  \begin{center}
    \FigureFile(80mm,80mm){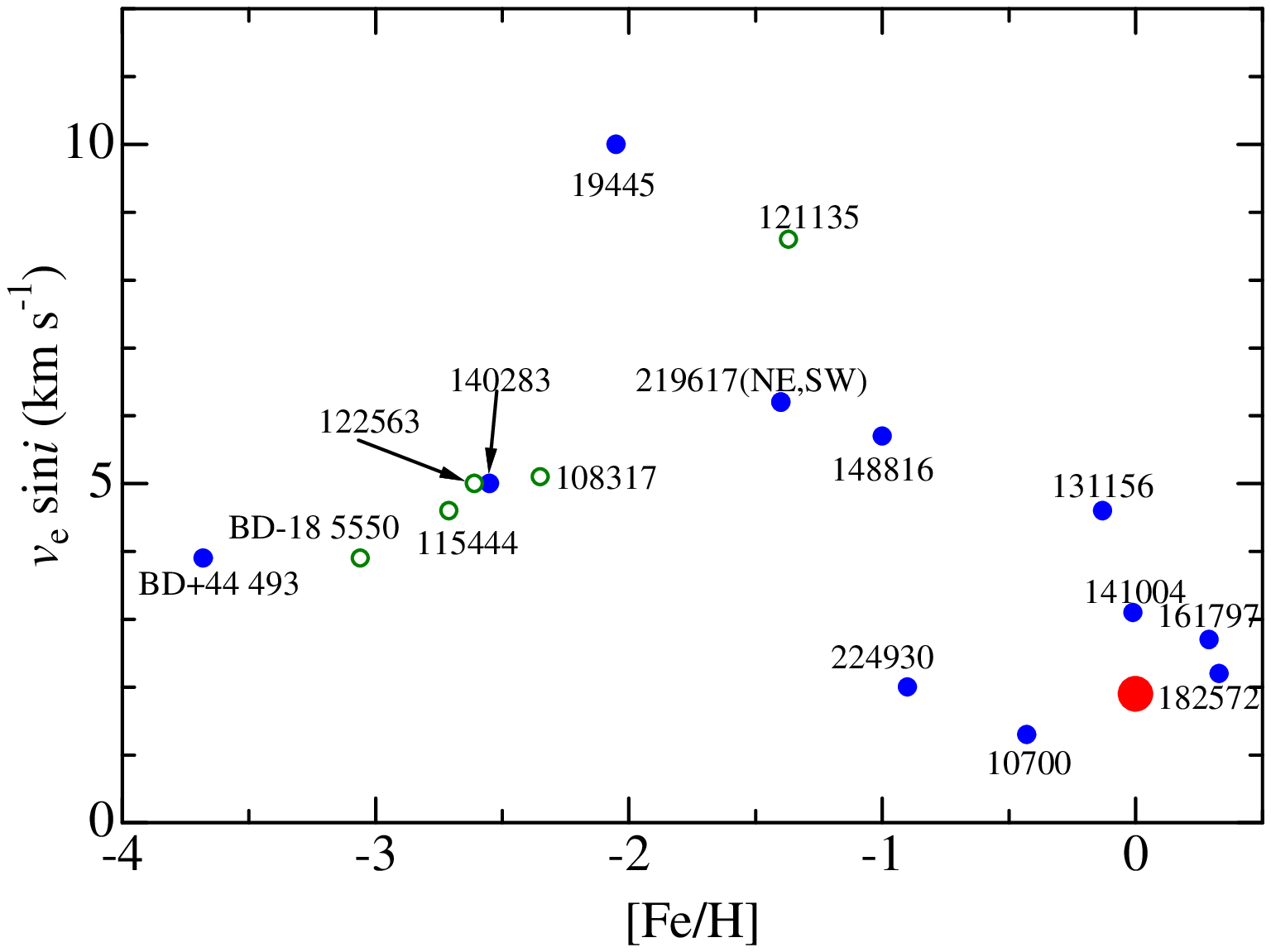}
  \end{center}
\caption{Published values of the projected rotational velocity 
($v_{\rm e}\sin i$) plotted against the metallicity ([Fe/H]) 
for 17 stars (among the 33 targets in this study) for which 
literature data are available.  
We consulted Cor\'{t}es et al.'s (2009) compilation for 11 metal-poor 
stars with [Fe/H]~$\le -1$, the Bright Star Catalog (Hoffleit \& Jaschek 1982) 
only for HD~224930 ([Fe/H]~=~$-0.9$), and Valenti and Fischer (2005) 
for 5 stars with $-0.5 <$~[Fe/H]. The corresponding star designations
are indicated in the figure. The meanings of the symbols are the 
same as in figure 3.
}
\end{figure}

\subsection{Impact of Chromospheres upon Spectral Line Formation}

Given that the existence of high-temperature layer ($T\sim 10^{4}$~K)
in the upper atmosphere of any solar-type stars (regardless of
the metallicity) has been established, we have some concern about
whether this effect may have any influence on the analysis of
stellar spectra (e.g., chemical abundance determinations), where 
conventional model atmospheres (with monotonically decreasing
temperatures outwards) are exclusively relied upon. 
Actually, the same situation applied even in the recent 
state-of-the-art 3D inhomogeneous hydrodynamical model atmospheres, 
where the averaged temperatures are nearly the same (for the solar 
metallicity) or even appreciably {\it lowered} (for the metal-poor 
condition) compared to the classical 1D model (see, e.g., figures 1 
and 2 in Asplund 2005), reflecting the fact that any physical 
process (e.g., dissipation of acoustic energy) leading to formation 
of chromospheres is not included in the current 3D models.
Thus, such a difference in the temperature structure may have some 
appreciable effect on the spectral line formation (especially for 
high-forming strong lines), though it is not easy to make a general 
prediction without any detailed calculation. For example, the core 
flux of the strong Ca~{\sc ii} lines (e.g., doublet lines at 
3934/3968~$\rm\AA$ or triplet lines at 8498/8542/8662~$\rm\AA$) 
tends to be raised by an increase in 
the surface temperature (cf. Appendix B in Takeda et al. 2010). 
On the contrary, the core of the high-excitation O~{\sc i} lines 
(e.g., triplet lines at 7771--5~$\rm\AA$  or 8446~$\rm\AA$) gets 
{\it deepened} (though slightly) by the chromospheric temperature rise, 
as shown by Takeda (1995).

Besides, even for ordinary (comparatively weak) spectral lines, 
which are deep-forming and appear to be practically unaffected
by any temperature structure in high optically-thin layers, a 
{\it non-local} effect of chromospheric radiation could be of
significance, since emission of photons followed by electron 
recombination is considerably enhanced in the chromosphere
reflecting that recombination rates are sensitive to the 
local electron temperature ($T_{\rm e}$) with a dependence of 
$\propto \exp(-h\nu/kT_{\rm e})$ (where Wien's approximation 
is hold; $h$: Planck constant, $k$: Boltzmann constant). 
This effect might influence the ionization equilibrium of any 
element, especially in metal-deficient atmospheres where such 
extra radiation of chromospheric origin has a higher possibility 
to reach the lower photosphere without being absorbed.

\section{Conclusion}

Based on the near-IR spectra of 33 late-type stars (9 giants 
and 24 dwarfs) in the wide metallicity range
($-3.7 \ltsim$~[Fe/H]~$\ltsim +0.3$),
which were obtained with IRCS+AO188 of the Subaru Telescope, 
we examined whether the He~{\sc i} 10830 line is visible and 
how its strength behaves itself in the metal-poor regime, 
in order to investigate the chromospheric activity in these stars.

We confirmed that this line is seen in absorption in almost all 
stars; i.e., not only metal-rich stars of disk population 
but also extremely metal-poor stars (down to [Fe/H]~$\simeq -3.7$).
This manifestly indicates that chromospheres ubiquitously 
exist in these FGK-type stars regardless of the metallicity or age. 

The equivalent width of this line turned out nearly constant 
at $\log EW({\rm m\AA}) \sim 1.5$ over the wide metallicity range of
$-3.7 \ltsim$~[Fe/H]~$\ltsim -0.5$ corresponding to old 
halo and thick-disk stars, while it tends to show a large scatter 
(many clustering around $\log EW({\rm m\AA}) \sim 2$ on the average)
for comparatively young thin-disk stars at $-0.5 \ltsim$~[Fe/H],
from which we may speculate that the origin of chromospheres 
is different between these two star groups.

For the latter group of thin-disk stars, still showing appreciable
rotational velocities, stellar rotation may play the decisive 
role in controlling the chromospheric activity, for which 
organized magnetic fields generated by rotation-induced dynamo 
would be responsible, as widely believed. The difference in the
rotational velocity would be the reason why $EW$s in this group
are diversified.

Meanwhile, the standard rotation/magnetism-induced activity 
would not be relevant any more for the former group of 
old halo or thick-disk stars, whose rotational velocity should 
have been considerably slowed down. Instead, a constant 
``basal''-level chromospheric activity would come into sight, 
for which energy dissipation of acoustic waves generated 
by convection may be a promising heating mechanism. 
Long-term observations monitoring the activity of old metal-poor 
stars would be useful to substantiate this hypothesis, since 
such stars would not show any activity variation. 

Yet, we should also keep in mind that this scenario is based on the 
widely believed hypothesis that stellar rotation generally gets 
slowed down with age as a result of the magnetic braking effect.
It is therefore important to check whether rotational velocities 
of old metal-poor stars tend to be slower than those of young stars.
Given the enormous difficulty of establishing $v_{\rm e}\sin i$
of considerably slow rotators, however, this would not be an easy task 
(even the inverse trend is suggested from the literature data
of the present sample stars, though its credibility is questionable).

The existence of chromospheres may influence the formation of
spectral lines. Since current analysis of stellar spectra
such as abundance determinations almost exclusively rely on 
conventional model atmospheres without any consideration of 
such a temperature rise, whether and how the difference in the 
temperature structure of the upper atmosphere makes impact 
on specific cases may be worth investigation.

\bigskip

We express our heartful thanks to Y. Minowa and T.-S. Pyo
for their kind advices and helpful support in preparing
as well as during the IRCS+AO188 observations.

One of the authors (M. T.-H.) is grateful for a financial support 
from a grant-in-aid for scientific research (C, No. 22540255) 
from the Japan Society for the Promotion of Science.

This research has made use of the SIMBAD database operated at
CDS, Strasbourg, France, as well as the data obtained from 
the High Energy Astrophysics Science Archive Research Center 
(HEASARC), provided by NASA's Goddard Space Flight Center.

\clearpage
\onecolumn

\setcounter{figure}{1}
\begin{figure}
  \begin{center}
    \FigureFile(150mm,200mm){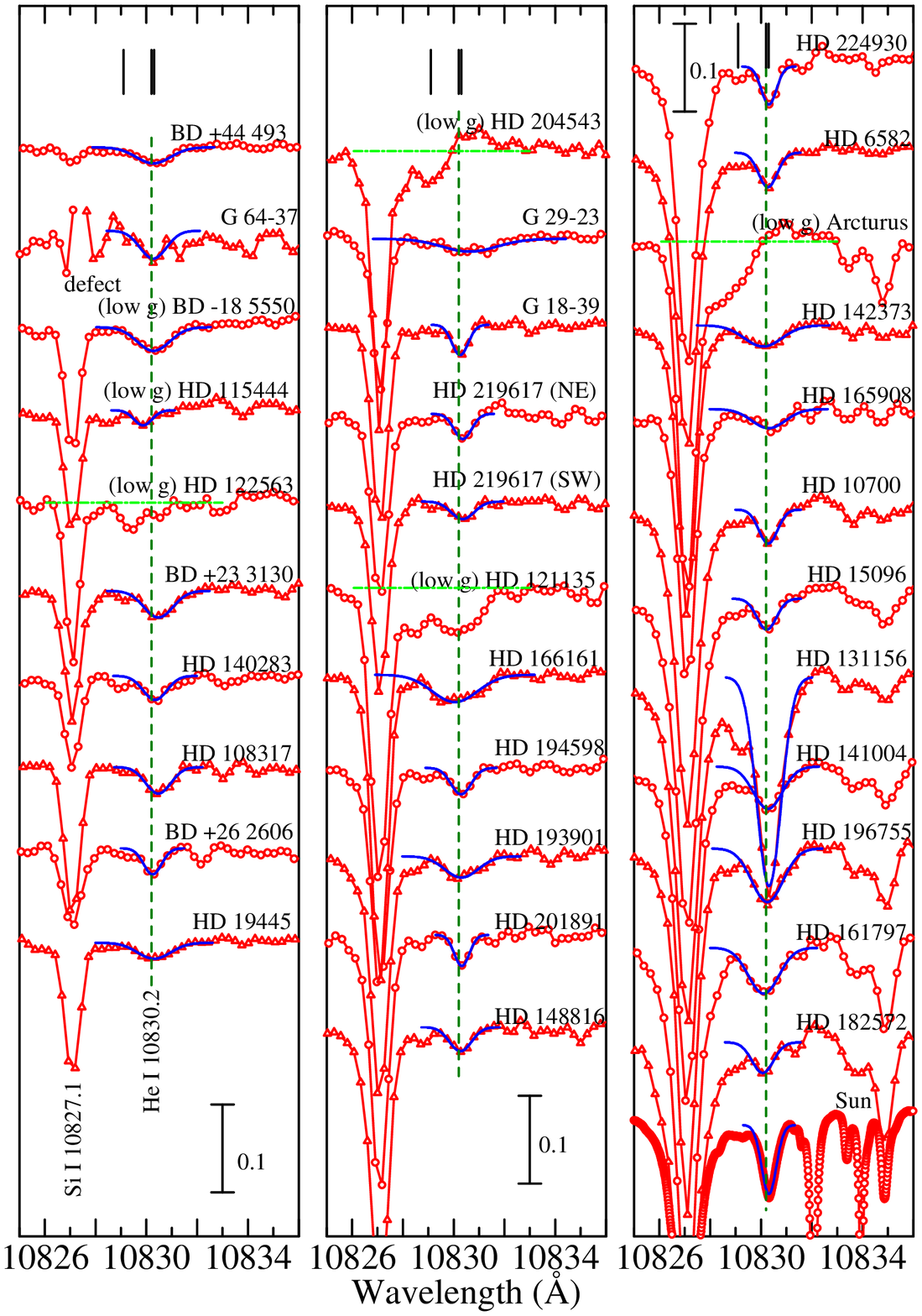}
  \end{center}
\caption{Observed spectra (red symbols) of 33 target stars (plus 
the Sun) in the 10825--10836~$\rm\AA$ region comprising the 
Si~{\sc i} 10827 and He~{\sc i} 10830 lines. The Gaussian-fit 
profiles applied to the He~{\sc i} 10830 line
for $EW$ measurements are depicted by blue lines.
The wavelength positions of three components (10829.088, 10830.247, 
and 10830.336~$\rm\AA$) are marked by ticks at the top of each panel,
while the strength-weighted mean wavelength of 10830.17~$\rm\AA$
(almost corresponding to the center of the He~{\sc i} 10830 line)
is shown by the vertical dashed line.
In each panel (from left to right), the spectra are arranged 
(from top to bottom) in the ascending order of [Fe/H] as in table 1. 
Each spectrum is shifted by 0.1 relative to the adjacent one. 
The wavelength scale of each stellar spectrum has been adjusted to the
rest frame. Stars indicated with ``(low $g$)'' at the head of the name
are ``low-gravity'' giants with $1 < \log g < 2$, for which measurements
tend to be difficult because they often show complex blue-shifted
components (blended with the Si~{\sc i} line) caused by substantial 
outflow. The Gaussian profiles fitted for $EW$ measurements are
depicted in blue lines, while only the continuum levels are indicated
by green dashed lines for the four ``low $g$'' cases where measurements 
could not be done.
}
\end{figure}

\clearpage
\setcounter{table}{0}
\begin{table}[h]
\caption{Program stars and measurement results for the He~{\sc i} 10830 line.}
\scriptsize
\begin{center}
\begin{tabular}{cc@{ }c@{ }c@{ }r r c cccl} 
\hline\hline
Name & $T_{\rm eff}$ & $\log g$ & $v_{\rm t}$ & [Fe/H] & $V$ & 
$f_{\rm x}/f_{\rm bol}$ & $R_{0}$ & $fwhm$ & $EW$ & Remark \\
 & (K) & (cm~s$^{-2}$) & (km~s$^{-1}$) & (dex) & (mag) & (dex) & & ($\rm\AA$) &(m$\rm\AA$)& \\
\hline
BD+44$^{\circ}$493   &  5510 &  3.70 &  1.30 & $-$3.68 & 9.13 & $\cdots$ &  0.018&   1.60 &  31.4 & D\\
G~64-37      &  6432 &  4.24 &  1.50 & $-$3.08 & 11.14 & $\cdots$ &  0.032&   1.23 &  41.7 & D, insufficient S/N, uncertain\\
BD$-$18$^{\circ}$5550  &  4750 &  1.40 &  1.80 & $-$3.06 & 9.35 & $\cdots$ &  0.027&   1.51 &  43.9 & G(lg)\\
HD~115444    &  4721 &  1.74 &  2.00 & $-$2.71 & 9.00 & $\cdots$ &  0.017&   0.84 &  15.5 & G(lg), profile not clear, uncertain\\
HD~122563    &  4572 &  1.36 &  2.90 & $-$2.61 & 6.20 & $\cdots$ &  $\cdots$& $\cdots$ &  $\cdots$ & G(lg), blue shift, abandoned\\
BD+23$^{\circ}$3130  &  5000 &  2.20 &  1.40 & $-$2.60 & 8.94 & $\cdots$ &  0.031&   1.35 &  44.3 & G\\
HD~140283    &  5830 &  3.67 &  1.90 & $-$2.55 & 7.21 & $\cdots$ &  0.028&   1.10 &  32.5 & D\\
HD~108317    &  5310 &  2.77 &  1.90 & $-$2.35 & 8.03 & $\cdots$ &  0.031&   1.16 &  38.0 & G\\
BD+26$^{\circ}$2606  &  5875 &  4.10 &  0.40 & $-$2.30 & 9.73 & $\cdots$ &  0.030&   0.83 &  26.5 & D\\
HD~19445     &  6130 &  4.39 &  2.10 & $-$2.05 & 8.05 & $\cdots$ &  0.018&   1.54 &  30.4 & D\\
HD~204543    &  4672 &  1.49 &  2.00 & $-$1.72 & 8.60 & $\cdots$ &  $\cdots$& $\cdots$ & $\cdots$ & G(lg), blue shift, abandoned\\
G~29-23      &  6194 &  4.04 &  1.50 & $-$1.69 & 10.19 & $\cdots$ &  0.015&   2.53 &  39.4 & D, profile not clear, uncertain\\
G~18-39      &  6093 &  4.19 &  1.50 & $-$1.46 & 10.38 & $\cdots$ &  0.035&   0.75 &  27.9 & D\\
HD~219617(NE)  &  5825 &  4.30 &  1.40 & $-$1.40 & $^{*}$8.16 & $\cdots$ &  0.029&   0.82 &  25.3 & D\\
HD~219617(SW)  &  5825 &  4.30 &  1.40 & $-$1.40 & $^{*}$8.16 & $\cdots$ &  0.020&   1.08 &  23.3 & D\\
HD~121135    &  4934 &  1.91 &  1.60 & $-$1.37 & 9.30 & $\cdots$ &  $\cdots$& $\cdots$ & $\cdots$ & G(lg), blueside blend, abandoned\\
HD~166161    &  5350 &  2.56 &  2.25 & $-$1.22 & 8.16 & $\cdots$ &  0.031&   2.10 &  69.9 & G\\
HD~194598    &  6020 &  4.30 &  1.40 & $-$1.15 & 8.36 & $\cdots$ &  0.031&   0.94 &  30.5 & D\\
HD~193901    &  5699 &  4.42 &  1.20 & $-$1.10 & 8.67 & $\cdots$ &  0.025&   1.55 &  40.5 & D\\
HD~201891    &  5900 &  4.19 &  1.40 & $-$1.10 & 7.38 & $\cdots$ &  0.035&   0.69 &  26.2 & D\\
HD~148816    &  5860 &  4.07 &  1.60 & $-$1.00 & 7.27 & $\cdots$ &  0.028&   1.03 &  30.1 & D\\
HD~224930    &  5275 &  4.10 &  1.05 & $-$0.90 & 5.75 & $-5.91$ &  0.044&   0.70 &  32.8 & D\\
HD~6582      &  5331 &  4.54 &  0.73 & $-$0.81 & 5.12 & $\cdots$ &  0.040&   0.86 &  36.7 & D\\
Arcturus    &  4281 &  1.72 &  1.49 & $-$0.55 & $-0.04$ & $\cdots$ & $\cdots$&  $\cdots$ &$\cdots$ & G(lg), blue shift, abandoned\\
HD~142373    &  5776 &  3.83 &  1.26 & $-$0.51 & 4.62 & $\cdots$ &  0.024&   1.74 &  45.1 & D\\
HD~165908    &  6183 &  4.35 &  1.24 & $-$0.46 & 5.07 & $-6.10$ &  0.022&   1.56 &  36.2 & D\\
HD~10700     &  5420 &  4.68 &  0.66 & $-$0.43 & 3.50 & $-6.33$ &  0.040&   0.87 &  36.4 & D\\
HD~15096     &  5375 &  4.30 &  0.80 & $-$0.20 & 7.95 & $\cdots$ &  0.036&   0.90 &  34.6 & D\\
HD~131156    &  5527 &  4.60 &  1.10 & $-$0.13 & 4.59 & $-4.47$ &  0.241&   1.13 & 291.0 & D, high-activity star (strong He line)\\
HD~141004    &  5877 &  4.11 &  1.17 & $-$0.01 & 4.43 & $-6.23$ &  0.049&   1.35 &  69.9 & D\\
HD~196755    &  5750 &  3.83 &  1.23 &  +0.09 & 5.05 & $-5.82$ &  0.061&   1.45 &  94.7 & D\\
HD~161797    &  5580 &  3.99 &  1.11 &  +0.29 & 3.42 & $-6.23$ &  0.053&   1.41 &  80.0 & D\\
HD~182572    &  5566 &  4.11 &  1.07 &  +0.33 & 5.16 & $-6.25$ &  0.036&   0.99 &  38.0 & D, blueside blend, rather uncertain\\
Sun         &  5780 &  4.44 &  1.00 &  0.00 & $\cdots$ & $\cdots$ &  0.080&   0.68 &  58.2 & D\\
\hline
\end{tabular}
\end{center}
In columns 1 through 7 are given the star designation, 
effective temperature, logarithmic surface gravity, 
microturbulent velocity dispersion, Fe abundance relative to
the Sun, apparent $V$ magnitude, and $\log (f_{\rm x}/f_{\rm bol})$
available in the ROSAT data archive (definitely identified cases).
See table 1 of Paper I for the source of atmospheric parameters.
Columns 8--10 present the results of the measurement for the 
He~{\sc i} 10830 line: $R_{0}$, $fwhm$, and $EW$ are the line-center depth 
($\equiv 1 - f_{0}/f_{\rm cont}$), the full-width at half-maximum, 
and the equivalent width derived from the Gaussian-fit measurement, 
respectively. The characters in column 10 denote the luminosity/gravity 
class of each star:``D'' is for dwarfs ($\log g >3$) and ``G'' is 
for giants ($\log g <3$); particularly low-gravity giants ($\log g <2$)
are further marked with ``(lg)'' (see also footnote 1).   
Some remarks relevant to specific cases are also given in column 10, 
when necessary. The objects are arranged in the ascending order of [Fe/H].\\
$^{*}$ Combined magnitude of the double-star system. 
\end{table}

\begin{thebibliography}{}
\bibitem[]{}
  Andretta, V., \& Giampapa, M. S. 1995, ApJ, 439, 405
\bibitem[]{}
  Asplund, M. 2005, ARA\&A, 43, 481
\bibitem[]{}
  Cayrel, R. 1988, in Proc. IAU Symp. 132, The Impact of Very High S/N
  Spectroscopy on Stellar Physics, ed. G. Cayrel de Strobel \& M. Spite
  (Dordrecht: Kluwer), 345
\bibitem[]{}
  Cayrel de Strobel, G., Soubiran, C., \& Ralite, N. 2001, A\&A, 373, 159
\bibitem[]{}
  Cor\'{t}es, C., Silva, J. R. P., Recio-Blanco, A., Catelan, M., 
  Do Nascimento, Jr., J. D., \& De Medeiros, J. R. 2009, ApJ, 704, 750
\bibitem[]{}
  Dupree, A. K., Hartmann, L. H., \& Smith, G. H. 1990, ApJ, 353, 623
\bibitem[]{}
  Dupree, A. K., Smith, G. H., \& Strader, J. 2009, AJ, 138, 1485
\bibitem[]{}
  Hoffleit, D., \& Jaschek, C. 1982, The Bright Star Catalogue, 4th ed.
  (New Haven: Yale University Observatory)
\bibitem[]{}
  Kurucz, R. L., Furenlid, I., Brault, J., \& Testerman, L. 1984,  
  Solar Flux Atlas from 296 to 1300 nm
  (Sunspot, New Mexico: National Solar Observatory)
\bibitem[]{}
  Lambert, D. L. 1987, ApJS, 65, 255
\bibitem[]{}
  Livingston, W., Wallace, L., White, O. R., \& Giampapa, M. S. 2007,
  ApJ, 657, 1137
\bibitem[]{}
  Narain, U., \& Ulmschneider, P. 1996, Space Sc. Rev., 75, 453
\bibitem[]{}
  Noyes, R. W., Hartmann, L. W., Baliunas, S. L., Duncan, D. K.,
  \& Vaughan, A. H. 1984, ApJ, 279, 763
\bibitem[]{}
  O'Brien, Jr., G. T., \& Lambert, D. L. 1986, ApJS, 62, 899
\bibitem[]{}
  Peterson, R. C., \& Schrijver, C. J. 1997, ApJ, 480, L47
\bibitem[]{}
  Peterson, R. C., \& Schrijver, C. J. 2001, 
  in The 11th Cool Stars, Stellar Systems and the Sun, 
  ASP Conf. Ser. Vol. 223, eds. R. J. Garc\'{\i}a L\'{o}pez, 
  R. Rebolo, and M. R. Zapatero (San Francisco: Astronomical
  Society of the Pacific), 300
\bibitem[]{}
  Robinson, R. D., Worden, S. P., \& Harvey, J. W. 1980, ApJ, 236, L155
\bibitem[]{}
  Rutten, R. G. M., Schrijver, C. J., Lemmens, A. F. P.,
  \& Zwaan, C. 1991, A\&A, 252, 203
\bibitem[]{}
  Schrijver, C. J. 1987, A\&A, 172, 111
\bibitem[]{}
  Schrijver, C. J. 1995, A\&AR, 6, 181
\bibitem[]{}
  Smith, G. H., \& Churchill, C. W. 1998, MNRAS, 297, 388
\bibitem[]{}
  Smith, G. H., \& Dupree, A. K. 1988, AJ, 95, 1547
\bibitem[]{}
  Takeda, Y. 1995, PASJ, 47, 463
\bibitem[]{}
  Takeda, Y., Honda, S., Kawanomoto, S., Ando, H., \& Sakurai, T. 2010,
  A\&A, 515, A93
\bibitem[]{}
  Takeda, Y., \& Takada-Hidai, M. 2011, PASJ, in press (Paper~I) 
  (Preprint: arXiv 1009.0824)
\bibitem[]{}
  Valenti, J. A., \& Fischer, D. A. 2005, ApJS, 159, 141
\bibitem[]{}
  Vaughan, Jr., A. H., \& Zirin, H. 1968, ApJ, 152, 123
\bibitem[]{}
  Zarro, D. M., \& Zirin, H. 1986, ApJ, 304, 365
\bibitem[]{}
  Zirin, H. 1982, ApJ, 260, 655

\end{thebibliography}
\end{document}